\pdfoutput=1 
\documentclass[11pt]{article}
\usepackage{graphicx,float,amsmath,amssymb,amsthm}
\usepackage{color}

\usepackage[english,francais]{babel}

\renewcommand{\em}{\it}

\newfont{\ensmathquatorze}{msbm10 scaled 1400}
\newfont{\ensmathonze}{msbm10 scaled 1100}
\newfont{\ensmathdix}{msbm10}
\newfont{\ensmathneuf}{msbm10 scaled 833}
\newfont{\ensmathhuit}{msbm10 scaled 694}
\newfam\ensmathfam                        
\textfont\ensmathfam=\ensmathonze        
\scriptfont\ensmathfam=\ensmathdix       
\scriptscriptfont\ensmathfam=\ensmathhuit
\def\ensmf{\fam\ensmathfam\ensmathonze}         

\renewcommand{\leq}{\leqslant}

\def\eqdef{\stackrel{\mbox{\tiny def}}{=}}     
\newcommand{\ket}[1]{|\kern.3ex#1\kern.3ex\rangle}
\newcommand{\bra}[1]{\langle\kern.3ex #1 \kern.3ex|}
\newcommand{\mean}[1]{\overline{ #1 }} 
\newcommand{\acos}{\mathop{\mathrm{arccos}}\nolimits}
\newcommand{\EXP}[1]{{\mbox{\large e}}^{#1}}         

\newcommand{\re}{\mathop{\mathrm{Re}}\nolimits}      

\def\NN{{\ensmf N}}                 
\def\ZZ{{\ensmf Z}}                 
\def\RR{{\ensmf R}}                 

\def\I{\mathrm{i}}                  
\def\D{\mathrm{d}}                  

\newcommand{\derivp}[2]{\frac{\partial #1}{\partial #2}}

\newcommand\antiddots{\mathinner{\mkern2mu\raise1pt\hbox{.}\mkern2mu
\newline \raise4pt\hbox{.}\mkern2mu\raise7pt\hbox{.}\mkern1mu}}

\newcommand{\diff}{\mathrm{d}}

\begin{document}

\selectlanguage{english}

\title{ Effect of boundaries on the spectrum of a \\  
  one-dimensional random mass Dirac Hamiltonian }

\author{ Christophe Texier$^{(a,b)}$ and Christian Hagendorf$^{(c)}$ }

\date{November 6, 2009}

\maketitle

\hspace{1cm}
$^{(a)}$
\begin{minipage}[t]{13cm}
{\small
 Laboratoire de Physique Th\'eorique et Mod\`eles Statistiques,
UMR 8626 du CNRS, 

Universit\'e Paris-Sud, B\^at. 100, F-91405 Orsay Cedex, France.
}
\end{minipage}

\vspace{0.25cm}

\hspace{1cm}
$^{(b)}$
\begin{minipage}[t]{13cm}
{\small
 Laboratoire de Physique des Solides, UMR 8502 du CNRS, 

Universit\'e Paris-Sud, B\^at. 510, F-91405 Orsay Cedex, France.
}
\end{minipage}

\vspace{0.25cm}

\hspace{1cm}
$^{(c)}$ 
\begin{minipage}[t]{13cm}
{\small
Laboratoire de Physique Th\'eorique de l'\'Ecole Normale Sup\'erieure,

24, rue Lhomond, F-75230 Paris Cedex 05, France.
}
\end{minipage}

\begin{abstract}
  The average density of states (DoS) of the one-dimensional Dirac
  Hamiltonian with a random mass on a finite interval $[0,L]$ is derived.
  Our method relies on the eigenvalues distributions (extreme value
  statistics problem) which are obtained explicitly. 
  The well-known Dyson singularity
  $\mean{\varrho(\epsilon;L)}\sim-\frac{L}{|\epsilon|\ln^{3}|\epsilon|}$
  is recovered above the crossover energy
  $\epsilon_c\sim\exp-\sqrt{L}$. 
  Below $\epsilon_c$ we find a log-normal suppression of the average DoS
  $\mean{\varrho(\epsilon;L)}\sim\frac1{|\epsilon|\sqrt{L}}\exp(-\frac1L\ln^2|\epsilon|)$.
\end{abstract}

\noindent
PACS numbers~: 72.15.Rn~; 73.20.Fz~; 02.50.-r.




\section{Introduction}
Let us consider the one-dimensional Dirac equation
$\mathcal{H}_\mathrm{D}\psi=\epsilon\psi$ for the Hamiltonian
\begin{equation}
  \label{HDirac}
  \mathcal{H}_\mathrm{D} = -\alpha_x\,\I\partial_x + \beta\, \phi(x)
  \:,
\end{equation}
acting on a two-component spinor.
$\phi(x)$ plays the role of a
mass and will be taken as random.
The Dirac matrices are chosen as $\beta=\gamma^0=\sigma_1$ and  
$\alpha_x=\gamma^0\gamma^1=-\sigma_2$, where $\sigma_i$ are the usual Pauli matrices,
corresponding to the representation of the Clifford algebra
$\gamma^0=\sigma_1$ and $\gamma^1=-\I\sigma_3$.
One-dimensional random Dirac Hamiltonians appear in several contexts
of condensed 
matter physics ranging from disordered half-filled metals \cite{GogMel77,Gog82},
random spin-chain models (random antiferromagnetic spin-$1/2$
chains~\cite{Fis94,McK96,BalFis97}, random transverse field Ising
spin-$1/2$ chains~\cite{Fis95,FisYou98,FisLeDMon98}, 
spin-Peierls chains and spin-ladders \cite{SheTsv98,SteFabGog98}),
or organic conductors~\cite{TakLinMak80}.  Another common
representation of Dirac 
matrices $\beta=\sigma_1$ and $\alpha_x=\sigma_3$, related to the one
chosen in the present article by a unitary transformation
$\mathcal{U}=(1+\I\sigma_1)/\sqrt2$, shows that the Dirac equation has
the form of the linearised Bogoliubov-de~Gennes equation describing a
superconductor with random gap~\cite{OvcEri77,LifGrePas88}.  
We obtain a connection to another important, and well-studied problem by squaring the Dirac Hamiltonian: 
$\mathcal{H}_\mathrm{D}^2=-\partial_x^2+\phi(x)^2+\I\gamma^1\phi'(x)=-\partial_x^2+\phi(x)^2+\sigma_3\phi'(x)$. 
This leads to a couple of isospectral supersymmetric Hamiltonians
$H_\pm=-\frac{\D^2}{\D{}x^2}+\phi(x)^2\pm\phi'(x)$. 
The corresponding Schr\"odinger equation can be transformed into a
Fokker-Planck equation 
$\partial_tP=\partial_x(\partial_x\mp2\phi(x))P$
describing classical diffusion in random force field,
studied in numerous 
works~\cite{Sin82,Gol84,Kes86,BouComGeoLeD90,AslPotSai90,FisLeDMon98,LeDMonFis99,MonLeD02}.  
If the random mass is a Gaussian white noise
with characteristics $\mean{\phi(x)}=\mu\,g$ and
$\mean{\phi(x)\phi(x')}-\mean{\phi}\,^2=g\,\delta(x-x')$, where
$\mean{\cdots}$ denotes the average with respect to its realisations,
an exact analytical expression for the density of states (DoS)
was obtained \cite{OvcEri77,BouComGeoLeD90} 
(see \S8.2 of Ref.~\cite{LifGrePas88})
\footnote{ Note also the Ref.~\cite{Boc99} where the most general 1d
  Dirac Hamiltonian $\mathcal{H}_\mathrm{D}=\sigma_2\,\I\frac{\D}{\D{}x}+\sigma_1\phi(x)+\sigma_3\lambda(x)+V(x)$
  was studied for $\phi$, $\lambda$ and $V$ Gaussian white noises. This study
  showed that there is no other DoS singularity than the one
  arising in the random mass case.  
}. 
Another solvable case is the one where the mass is chosen as a
telegraph noise~\cite{ComDesMon95} ({\it i.e.} with exponentially
decaying correlations). 
The same low energy properties are obtained.
This is related to the fact that any short-range correlated noise,
upon large scale renormalization, reduces to a Gaussian white noise.

Here we focus on the case of Gaussian white noise with $\mean{\phi(x)}=0$
which is known to yield a Dyson singularity for the DoS
$\varrho(\epsilon\to0)\simeq\frac{g}{|\epsilon|\ln^{3}(g/|\epsilon|)}$
and a delocalisation transition at~$\epsilon\to0$. 
The aim of the present article is to study how the Dyson singularity
of the DoS is affected by boundary conditions.
We denote by  $\varrho(\epsilon;L)$ the DoS of the Hamiltonian
\eqref{HDirac} on a finite interval $[0,L]$.  
We will obtain the behaviour of the average DoS
$\overline{\varrho(\epsilon;L)}$ for a finite length.

\section{Boundary conditions}
Let us specify the two different types of boundary conditions we shall consider.

\vspace{0.25cm}
\noindent{\it Type (D)~:}
A simple way to introduce boundary conditions ensuring confinement
in a domain $\mathcal{D}$ is to consider the Dirac equation
$[\I\partial\hspace{-0.2cm}/-\phi(x)]\psi(x,t)=0$ with infinite mass
$\phi(x)=\Phi\to\infty$ for $x$ outside $\mathcal D$. This leads
  to the so-called ``bag model'' of hadronic physics
  \cite{ChoJafJohThoWei74}~: the boundary conditions are
  $(1+\I\vec{n}\cdot\vec\gamma)\psi|_{\partial\mathcal{D}}=0$, where
  $\vec{n}$ is the unit vector normal at the boundary
  $\partial\mathcal{D}$ of the domain. They force the vanishing of the
  component of the current density
  $\vec{\mathcal{J}}_\psi=\bar\psi\vec\gamma\psi$ perpendicular to the
  boundary, where $\bar\psi=\psi^\dagger\gamma^0$. 
In our case, for $x\in\RR^-$ the stationary solution of
the Dirac equation for a constant mass is a plane wave
$\psi(x)=u\,\EXP{\I{}kx}$ where the spinor $u$ is solution of
$(\alpha_xk+\beta\Phi)u=\epsilon{}u$, for $\epsilon^2=k^2+\Phi^2$. 
In the limit $\Phi\to\infty$ it becomes an evanescent wave with
$k\simeq-\I\Phi$~; this yields $(1-\I\gamma^1)\psi(0)=0$.
Similarly, at the other boundary we find $(1+\I\gamma^1)\psi(L)=0$.
We denote these constraints as ``type (D) boundary conditions''.
With our representation of the Clifford algebra, they read
$(1-\sigma_3)\psi(0)=(1+\sigma_3)\psi(L)=0$, {\it i.e.} each of the two
components of the bispinor $\psi=(\varphi,\chi)$ must vanish at one
side~: $\chi(0)=\varphi(L)=0$.
This condition obviously ensures the absence of Dirac current flow
$\mathcal{J}_\psi=-\psi^\dagger\sigma_2\psi$ across the boundaries.

\vspace{0.25cm}
\noindent{\it Type (S)~:}
A second interesting choice is given by
$(1+\sigma_3)\psi(0)=(1+\sigma_3)\psi(L)=0$. These conditions
coincide with Dirichlet boundary conditions for the associated
supersymmetric Schr\"odinger Hamiltonian~$H_\pm$ and will be denoted as
``type~(S)''\footnote{
  \noindent{\it General boundary conditions~:}
  It is possible to set up more general conditions forcing the absence
  of probability current flow at the boundaries~: these are the set of
  conditions parametrised by some real number $\vartheta$~\cite{AntComKne90}~:
  $(\EXP{\I\vartheta\gamma^5}+\I\vec{n}\cdot\vec\gamma)\psi|_{\partial\mathcal{D}}=0$
  where $\gamma^5$ is the chirality matrix, given by
  $\gamma^5=\gamma^0\gamma^1$ in dimension $d=1+1$.
  These general boundary conditions are obtained as follows~:
  one considers the Dirac equation in a bounded domain
  $\mathcal{D}$. The boundary conditions can be written
  $\I\vec{n}\cdot\vec\gamma\psi|_{\partial\mathcal{D}}=(A+B\,\gamma^5)\psi|_{\partial\mathcal{D}}$,
  where $\vec{n}$ is the unit vector normal to the domain, $A$ and $B$
  two complex numbers.
  The fact that $(\vec{n}\cdot\vec\gamma)^2=-1$ leads to $A^2-B^2=1$.
  Upon imposing the normal component of the Dirac current to vanish
  $\vec{n}\cdot\vec{\mathcal{J}}_\psi|_{\partial\mathcal{D}}=0$, we find
  $A\in\RR$ and~$B\in\I\RR$, and thus the desired form.
  With the representation chosen in our
  article these general boundary conditions take the form
  $(\EXP{-\I\vartheta_\pm\sigma_2}\pm\sigma_3)\psi(x_\pm)=0$ with $x_-=0$ and $x_+=L$. For type (D) we have $\vartheta_+=\vartheta_-=0$, and for (S) $\vartheta_-=\pi,\,\vartheta_+=0$.  Since $\vartheta_\pm\neq0$ or $\pi$ breaks the
  particle-hole symmetry, this case will not be considered here.
}.

\section{Eigenvalue distributions}
We now consider the Dirac Hamiltonian (\ref{HDirac}) on $[0,L]$ for
the two kinds of boundary conditions introduced above. The particle-hole symmetry
takes the form $\psi\to\tilde\psi=\sigma_3\psi$~:
$\sigma_3\mathcal{H}_\mathrm{D}\sigma_3=-\mathcal{H}_\mathrm{D}$.
Since both boundary conditions preserve the particle-hole symmetry, eigenvalues
appear in pairs $\pm\epsilon_n$ (by convention we choose $n=1,\,2,\,3,\cdots$
and~$\epsilon_n>0$).  We denote by
$\varrho(\epsilon)$ the DoS per unit length for an {\it infinite}
volume. It is related to the DoS of the finite-size system by 
$\varrho(\epsilon)=\lim_{L\to\infty}\varrho(\epsilon;L)/L$.
Because of particle-hole symmetry we can restrict ourselves to $\epsilon>0$.

Our method to obtain the average DoS $\mean{\varrho(\epsilon;L)}$
relies on the evaluation of the eigenvalue distributions
$\mathcal{W}_n(\epsilon;L)\eqdef\mean{\delta(\epsilon-\epsilon_n[\phi,L])}$,
derived in Ref.~\cite{Tex00} for (S)-boundaries.
Finding the distributions of the (ordered) variables
$\epsilon_n$ corresponds to an ``extreme value problem'' (here for
correlated variables). 
We recall the idea of the method~:
let us imagine that we impose the boundary conditions to hold solely at one end of the interval, say $x=0$. 
As $x$ increases the
two components of the spinor $\psi(x)=(\varphi(x),\chi(x))$, which solves the
Dirac equation, vanish alternately~; we denote by $\Lambda_m$ the
distance between a node of $\varphi(x)$ and the closest node of $\chi(x)$.
Since the evolution of $\psi(x)$ is Markovian, the $\Lambda_m$'s are
identical and independently distributed ({\it i.i.d.}) random variables whose
statistical properties are obtained by solving a first passage time
problem (see Ref.~\cite{Tex00} for details). 
Let us introduce the distribution $\pi_M(y)$ of the sum of $M$ {\it i.i.d.} 
rescaled lengths
$y=(\Lambda_1+\cdots+\Lambda_M)\mathcal{N}(\epsilon)$,
where
\begin{equation}
  \mathcal{N}(\epsilon)=\int_0^\epsilon\D\epsilon'\,\varrho(\epsilon')
  \underset{\epsilon\to0}{=}
  \frac{g/2}{[\ln(2g/\epsilon)-\mathrm{C}]^2+\pi^2/4}+
  O\left(\frac{\epsilon^2}{\ln^2\epsilon}\right)
  \approx \frac{g}{2\ln^2(g /\epsilon)}  \label{eq:IDoS}
\end{equation}
is the integrated DoS per unit length for an infinite volume~;
$\mathrm{C}=0.577...$ is the Euler-Mascheroni constant (the exact
expression for $\mathcal{N}(\epsilon)$ may be found in
Refs.~\cite{OvcEri77,LifGrePas88,BouComGeoLeD90}). This rescaling is
motivated by the fact that 
$\mean{\Lambda_m}=1/(2\mathcal{N}(\epsilon))$ \cite{Tex00}, and hence
$\mean{y}=M/2$.    
Both boundary conditions at $x=0,L$ are satisfied whenever the sum 
$\Lambda_1+\cdots+\Lambda_M$ coincides with the length $L$.
Therefore the distribution $\mathcal{W}_n(\epsilon;L)$ is
given by~\footnote{
  Note that the change of variable from $\epsilon$ to
  $L\,\mathcal{N}(\epsilon)$ corresponds to the so-called spectrum
  unfolding leading to a unit density of variables.}
\begin{equation}
  \mathcal{W}_n(\epsilon;L)
  = L\, \varrho(\epsilon)
  \ \varpi_n\Big( L\, \mathcal{N}(\epsilon) \Big)
  \:,
\end{equation}
where the dimensionless function $\varpi_n(y)$ is related to the
distributions $\pi_M(y)$ as follows~:

\vspace{0.25cm}

\noindent{\it Type (S)~:}
The same component of the Dirac spinor must vanish at the two sides of
the interval. Therefore $L$ must coincide with a sum of an {\it even} number
of lengths $\Lambda_m$'s (see eq.~(132) of Ref.~\cite{Tex00})~:
\begin{equation}
  \label{varpiSn}
  \boxed{  \varpi_n^{(S)}(y) =  \pi_{2n}(y) }
  \:.
\end{equation}

\vspace{0.25cm}

\noindent{\it Type (D)~:}
Both spinor components must vanish at one side of the interval.  Hence $L$
must coincide with a sum of an {\it odd} number of lengths $\Lambda_m$'s~:
\begin{equation}
  \label{varpiDn}
  \boxed{   \varpi_n^{(D)}(y) = \pi_{2n-1}(y) }
  \:.
\end{equation}

\vspace{0.25cm}

The distributions $\pi_n(y)$ are explicitly known in the low energy
limit $\epsilon \ll g$, {\it i.e.} for $gL \gg 1$ where
$\mathcal{W}_n(\epsilon;L)$ is essentially concentrated below $g$. 
Indeed, the characteristic
function of the lengths $\Lambda_m$'s is given by~\cite{Tex00}
$\mean{\EXP{-\alpha\Lambda}}\simeq\cosh^{-1}\sqrt{\alpha/\mathcal{N}(\epsilon)}$.
Whence we may write   
\begin{equation}
  \label{DefPin}
  \pi_n(y) \simeq \int_\mathcal{B}\frac{\D q}{2\I\pi}\,
  \frac{\EXP{q\,y}}{\cosh^{n}\sqrt{q}}
  \:
\end{equation}
where the integration is taken along the Bromwich contour.
Starting from this integral representation we obtain the following explicit formulae
\begin{subequations}
\begin{equation}
  \label{eqn:piodd}
  \boxed{
  \pi_{2k+1}(y) = \frac{2^{2k}}{\sqrt\pi\,y^{3/2}}
   \sum_{n=0}^\infty(-1)^{n+k}(2n+1)
   \left(
   \begin{array}{c}
   	  n+k\\
	  n-k
   \end{array}
   \right)
   \,\EXP{-\frac{(n+1/2)^2}y}
  }
\end{equation}
\begin{equation}
  \label{eqn:pieven}
  \boxed{
	\pi_{2k}(y) = \frac{2^{2k}}{\sqrt{\pi}\,y^{3/2}}\sum_{n=0}^\infty(-1)^{n+k}n
	\left(
	\begin{array}{c}
		 n+k-1\\
		n-k
	\end{array}
	\right)\EXP{-n^2/y}
  }\:
\end{equation}
\end{subequations}
whose demonstration is provided in appendix~\ref{app:pin}.

\section{DoS on a finite interval}
The average DoS can be related to the distributions recalled above~:
\begin{equation}
  \mean{\varrho(\epsilon;L)} = \sum_{n=1}^\infty \mathcal{W}_n(\epsilon;L)
  \:.
\end{equation}
Summation over $n$  leads to
\begin{equation}
  \label{RES1}
  \boxed{
  \mean{\varrho(\epsilon;L)}  = L\, \varrho(\epsilon)
  \ \mathcal{D}\Big( L\, \mathcal{N}(\epsilon) \Big)
  }
\end{equation}
where the dimensionless function $\mathcal{D}(y)$ depends on boundary
conditions. 
Note that this DoS has some interest for studying the problem
of 1d classical diffusion in a random force field with dilute
absorbers~\cite{TexHag09} (see also Ref.~\cite{LeD09} for an analysis
of this problem with the real space renormalisation group).

\vspace{0.25cm}

\noindent{\it Boundary conditions of type (S).--} 
The summation of the (S)-type distributions (\ref{varpiSn}) gives
\begin{equation}
  \mathcal{D}_S(y) = \int_\mathcal{B} \frac{\D q}{2\I\pi}\,
  \frac{\EXP{q\,y}}{\sinh^{2}\sqrt{q}}
  \:.
\end{equation}
The integrand is meromorphic in the complex plane 
(there is no branch
cut since $\sqrt{q}$ is the argument of an even function).  
The integral can be computed from the residue theorem.  The integrand
possesses a single pole at $q=0$ with residue equal to unity and an
infinite number of double poles on the real axis at $q=q_n=-(n\pi)^2$,
with $n=1,\,2,\,3,\cdots$. Using that
$\sinh^{2}\sqrt{q}\underset{q\sim q_n}{\simeq}\frac1{4q_n}{(q-q_n)^2}$,
we find the residues 
$
  \mathrm{Res}\big[\frac{\EXP{qy}}{\sinh^{2}\sqrt{q}}\,;\,q_n\big]
  = \frac{\D}{\D q}
  \big[\frac{(q-q_n)^2\EXP{qy}}{\sinh^{2}\sqrt{q}}\big]_{q=q_n}
  =(2+4q_n)\,\EXP{q_ny}
$.
Therefore  we have
\begin{equation}
  \label{Da}
  \mathcal{D}_S(y) = 
  1 +2\sum_{n=1}^\infty\left(1-2(n\pi)^2y\right)\,\EXP{-(n\pi)^2y}
  = \frac{4}{\sqrt{\pi}\,y^{3/2}}
  \sum_{n=1}^\infty n^2\EXP{-n^2/y}.
\end{equation}
where the second series expansion may readily be found from Poisson's
summation formula (see appendix~\ref{app:poisson}). 
We may also check that the summation of \eqref{eqn:pieven} leads to~\eqref{Da}.

\vspace{0.25cm}

\noindent{\it Boundary conditions of type (D).--} 
Compared to the (S) case, an additional $\cosh\sqrt{q}$ appears in the
integrand upon summation of \eqref{varpiDn}~:
\begin{equation}
  \mathcal{D}_D(y) = \int_\mathcal{B} \frac{\D q}{2\I\pi}\,
  \frac{\cosh\sqrt{q}}{\sinh^{2}\sqrt{q}}\,\EXP{q\,y}
  \:.
\end{equation}
It adds a sign $(-1)^n$ to the residues obtained
in the previous case and we thus find~
\begin{equation}
  \label{DaD}
  \mathcal{D}_D(y) = 
  1 +2\sum_{n=1}^\infty (-1)^n\left(1-2(n\pi)^2y\right)\,\EXP{-(n\pi)^2y}= \frac{1}{\sqrt{\pi}\,y^{3/2}}
  \sum_{n=0}^\infty (2n+1)^2\EXP{-(2n+1)^2/4y}\:.
\end{equation}

\vspace{0.25cm}

\begin{figure}[!ht]
  \centering
  \includegraphics[scale=0.85]{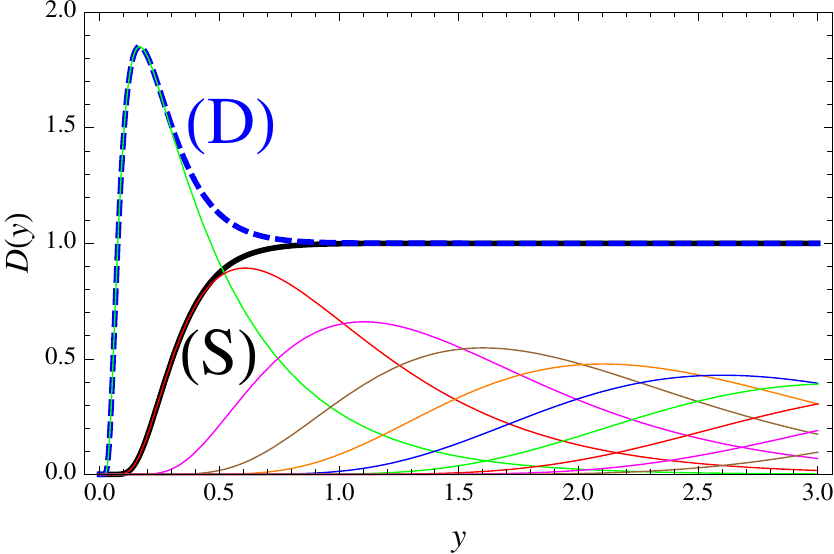}
  \caption{\it Functions $\mathcal{D}_S(y)$ (black continuous line)
    and $\mathcal{D}_D(y)$ (dashed blue line).
    First distributions $\varpi_n(y)$ are also plotted in thin lines.} 
  \label{fig:DSandDD}
\end{figure}

The two different boundary conditions can be treated on the same
footing by writing 
\begin{align}
  \label{Dsamefoot}
  \mathcal{D}_{S,D}(y) 
  = \left(1+2y\frac{\D}{\D y}\right)\sum_{n\in\ZZ}(\pm1)^n\EXP{-(n\pi)^2y}
\end{align}
and
\begin{align}
  \label{RES2}
  \boxed{
  \mathcal{D}_{S,D}(y) 
  = \frac{4}{\sqrt{\pi}\,y^{3/2}}
  \sum_{n=0}^\infty \left(n+\eta\right)^2\EXP{-\frac{(n+\eta)^2}y}
  }
  \quad
     \mbox{with }
     \eta=
     \left\{
     \begin{array}{ll}
     0   &\mbox{ for } S \\ [0.15cm]
     1/2 &\mbox{ for } D 
     \end{array}
     \right.
  \:.
\end{align}
This second series expansion allows to analyse the low energy
behaviour. 
It is worth noticing that this representation for
$\mathcal{D}(y)$ is very similar to the
corresponding one for $\varpi_1(y)$, for both kind of boundary conditions. 
In the (S) case eqs.~\eqref{pi2b} and {\it r.h.s.} of \eqref{DaD}
only differ by the $(-1)^{n+1}$ in the sum, while in the (D) case 
\eqref{Da} and {\it r.h.s.} of \eqref{pi1b} differ by a $(-1)^{n}(2n+1)$.
As a result we have~$\mathcal{D}(y\to0)\simeq\varpi_1(y)$ in both
cases. 
The two functions are plotted on figure~\ref{fig:DSandDD}. 
$\mathcal{D}_{S}(y)$ presents a monotonous behaviour what 
leads to a diminution of the low energy DoS for small energies.
Interestingly for (D) boundary conditions, while exponentially
suppressed for $y\to0$, $\mathcal{D}_{D}(y)$ increases for
intermediate values of $y$~; as a result (D) boundary conditions
induce an increase of the DoS at intermediate energies.

The low energy DoS presents a log-normal suppression~:
\begin{equation}
  \label{DosLimS}
  \boxed{
  \mean{ \varrho_S(\epsilon;L) }
  \simeq \frac{16}{|\epsilon|\,\sqrt{2\pi gL}}\,
  \EXP{-\frac2{gL}\ln^2(g/|\epsilon|)}
  \hspace{0.5cm}\mbox{for}\hspace{0.5cm} 
          |\epsilon| \ll \epsilon_c=g\,\EXP{-\sqrt{gL/2}} 
  }
\end{equation}
and 
\begin{equation}
  \label{DosLimD}
  \boxed{
  \mean{ \varrho_D(\epsilon;L) }
  \simeq \frac{4}{|\epsilon|\,\sqrt{2\pi gL}}\,
  \EXP{-\frac1{2gL}\ln^2(g/|\epsilon|)}
  \hspace{0.5cm}\mbox{for}\hspace{0.5cm} 
          |\epsilon| \ll \epsilon'_c=g\,\EXP{-\sqrt{2gL}} 
  }
\end{equation}
The DoS reaches its maximal value at $\epsilon_*\approx{}g\EXP{-gL/4}$
in the (S) case and at $\epsilon'_*\approx{}g\EXP{-gL}$ in the (D) case.

The weaker log-normal suppression in the (D) case as compared to the (S) case
is due to the fact that for a given realisation of the disorder, the
spectra for the two different kinds of boundary conditions are
such that 
$\epsilon_1^D<\epsilon_1^S<\epsilon_2^D<\epsilon_2^S<\cdots$, where
$\{\epsilon^{D,S}_n\}$ denotes the spectrum for boundary conditions of
type (D) and~(S) respectively.

Since $\mathcal{D}(y\to\infty)\simeq1$, eqs.~(\ref{Da},\ref{DaD}),
we recover the Dyson singularity for intermediate energies, as expected~:
\begin{equation}
  \mean{\varrho(\epsilon;L)} 
  \simeq L\:\varrho(\epsilon) 
  \approx \frac{L\,g}{|\epsilon|\ln^3(g/|\epsilon|)} 
  \hspace{.5cm}\mbox{for}\hspace{0.5cm} 
  \epsilon_c   \ll |\epsilon| \ll g
  \:.
\end{equation}
At higher energies $|\epsilon|\gg{}g$ one should recover the free
DoS~: $\mean{\varrho(\epsilon;L)}\simeq{L}/\pi$.

\begin{figure}[!ht]
  \centering
  \includegraphics[scale=0.9]{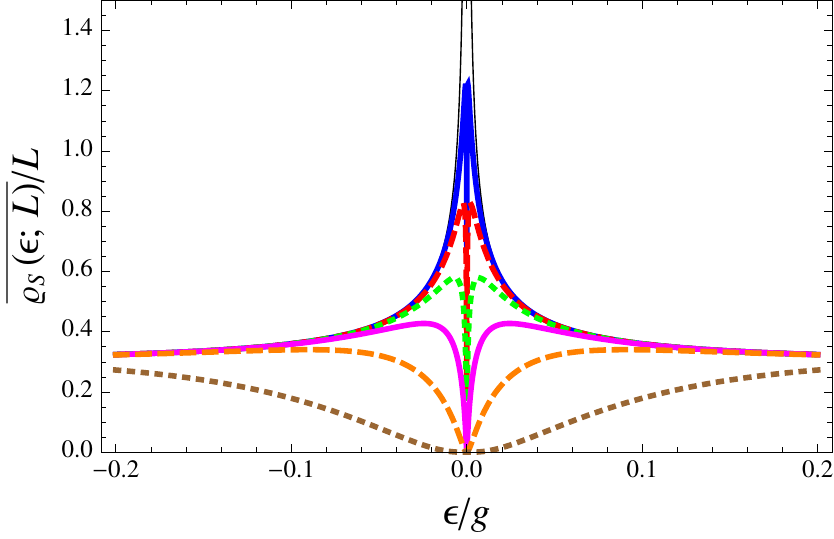}
  \hspace{0.5cm}
  \includegraphics[scale=0.9]{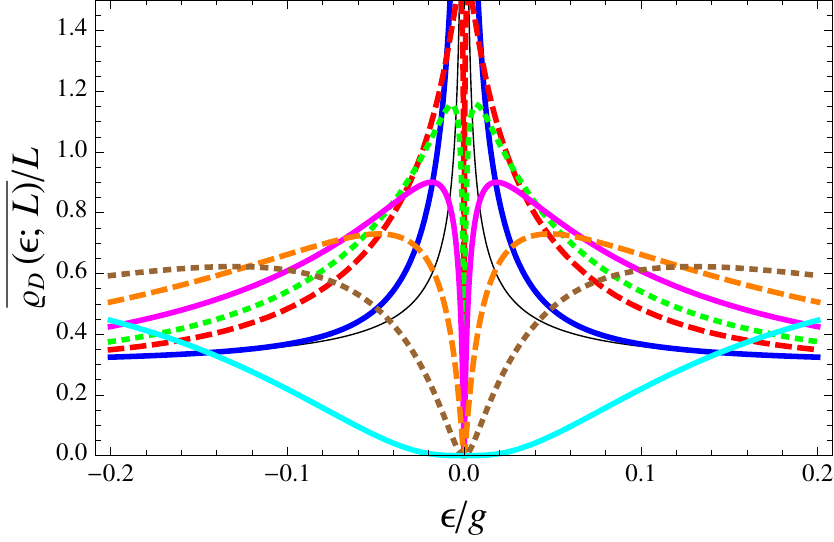}
    \caption{{\it Average DoS per unit length for a finite $L$~; 
    parameter is $g=1$.}
    Left~: {\it (S) boundary conditions
    for sizes $L=\infty$, $30$, $25$, $20$, $15$, $10$ \& $5$.}
    Right~: {\it (D) boundary conditions
    for sizes $L=\infty$, $12$, $6$, $5$, $4$, $3$, $2$ \& $1$.}
    }
  \label{fig:dos}
\end{figure}

The average DoS $\mean{\varrho(\epsilon;L)}$ is represented for various
values of $L$ on figure~\ref{fig:dos}. 
For the lowest energies $\epsilon\lesssim\epsilon_*\sim{}g\EXP{-gL}$
the DoS is suppressed in both cases. In the intermediate range
$\epsilon_*\lesssim\epsilon\lesssim\epsilon_c\sim{}g\EXP{-\sqrt{gL}}$,
the effect of the boundaries is to reduce the DoS in the (S) case 
but surprisingly to increase the DoS in the (D) case.

As $L$ is decreased the Dyson singularity is rapidly converted 
to a strong depletion of the low energy DoS.
For (S) boundary conditions this occurs for a surprisingly
relatively large length $L_*\sim10/g$. 
For the (D) case the memory of the Dyson singularity persists up to
smaller lengths since an increase of the low energy DoS is
apparent up to~$L'_*\sim1/g$ (figure~\ref{fig:dos}).  

\section{Conclusion}
By using known results on the energy-levels distributions
$\mathcal{W}_n(\epsilon;L)$ we have obtained the average density of states of
a random Dirac Hamiltonian on a bounded domain $[0,L]$.
For $gL\gg1$, we have seen that the DoS does not feel the boundary as
energies become larger than the energy scale
$\epsilon_c\sim{}g\,\EXP{-\sqrt{gL}}$.
In the intermediate range
$\epsilon_*\lesssim\epsilon\lesssim\epsilon_c$,
where $\epsilon_*\sim{}g\,\EXP{-gL}$, the DoS may be diminished by
boundaries, (S) case, or increased, (D) case.
For the lowest energies the DoS presents a log-normal suppression for
both kind of boundary conditions.

Note that the effect of one boundary condition was already studied in
Ref.~\cite{SteFabGog98} where the average local DoS on a semi-infinite
line was computed by the Berezinsk{\u\i}i technique.
These authors found an {\it increase} of the local DoS
by a factor $\ln(g/|\epsilon|)\gg1$ close to the boundary (at a
distance much smaller than the scale 
$g^{-1}\ln^2(g/|\epsilon|)$).
Here, by imposing boundary conditions at the two sides of finite interval,
we have shown that boundary conditions can induce both an increase or
a decrease of the DoS.

It is interesting to interpret our result within the classical
diffusion problem. The (S) boundary corresponds to diffusion in a
random force field with two absorbing boundaries while the (D) case
corresponds to one reflecting boundary and one absorbing.
The return probability is the Laplace transform of the DoS, therefore
the log-normal suppression of the average DoS
coincides to a log-normal decay of the average return probability
$\int_0^L\D{}x\,\overline{P(x,t\to\infty|x,0)}\sim\exp-\frac1{gL}\ln^2(g^2t)$.

To close this paper, let us examine the relation between our main
result (\ref{RES1},\ref{RES2}) with the localisation properties. 
It is only below the crossover energy
$\epsilon_c\sim{}g\,\EXP{-\sqrt{gL}}$ that eigenstates do feel the 
boundaries. The transition corresponds to the case where the
localisation length is of order of the system size
$\xi_{\epsilon_c}\sim{}L$, therefore this criterion suggests that
the eigenstate of energy $\epsilon$ is localised on a
scale~$\xi_\epsilon\sim g^{-1}\ln^2(g/|\epsilon|)$. 

The question of localisation was studied in several
works (Refs.~\cite{EggRie78,Zim82} for a tight binding Hamiltonian
with random hoppings and in Ref.~\cite{BouComGeoLeD90}
for the continuous supersymmetric Hamiltonian) that have demonstrated
the vanishing of the Lyapunov exponent~\footnote{We recall that the
  Lyapunov exponent characterises the exponential growth rate of the
  envelope of the wave function \cite{AntPasSly81,LifGrePas88}
  $\gamma\eqdef\lim_{x\to\infty}\frac{\Xi(x)}{x}$ where $\Xi$ controls
  the envelope of the wave function
  $\psi(x)=\EXP{\Xi(x)}\times$oscillations. } 
at the band edge $\gamma(\epsilon\to0)\sim{}g\,\ln^{-1}(g/|\epsilon|)$.
The Lyapunov exponent provides another possible definition of the
inverse localisation length $\widetilde\xi_\epsilon\eqdef1/\gamma$,
therefore much shorter than the previous one~:
$\widetilde\xi_\epsilon\sim g^{-1}\ln(g/|\epsilon|)\ll\xi_\epsilon$. 
In other terms the Lyapunov exponent analysis would put the
``localisation threshold''\footnote{
  Note that in the scaling theory of localisation \cite{KraMac93} the
  notion of a localisation threshold (or mobility edge)
  designates the energy separating
  localised and delocalised states for the {\it infinite} system.
  Within this usual terminology the model we are studying is always in
  a localised phase apart strictly at $\epsilon=0$.
  What we denote here by ``localisation threshold'' is the energy
  separating localised and delocalised states for a {\it finite}
  system size $L$. 
  With this definition there can not be a sharp localisation
  transition, due to fluctuations over disorder realisations.} at 
$\epsilon_*\sim{}g\EXP{-gL}$, much below
than~$\epsilon_c\sim{}g\,\EXP{-\sqrt{gL}}$.  

The existence of two characteristic length scales was pointed out in
several works~\cite{BouComGeoLeD90,Fis94,BalFis97,SteFabGog98}.  
These authors have found that the average Green's function 
({\it i.e.} the two point correlation function)
decays over~\footnote{Note that a related quantity first
  obtained Refs.~\cite{Gol84,Kes86} is  
  the propagator, the inverse Laplace transform of the Green's function,
  or more precisely the average conditional probability 
  $\mean{P(x,t|x',0)}=\mean{\phi_0(x)\phi_0(x')^{-1}\bra{x}\EXP{-tH_+}\ket{x'}}$
  of the related Fokker-Planck equation,
  where $\phi_0(x)$ is the zero mode of the supersymmetric Hamiltonian $H_+$~; the result of
  these works was reproduced in Ref.~\cite{ComDea98} from stochastic
  Riccati analysis and in Ref.~\cite{LeDMonFis99} with the real space
  renormalisation group.} 
$\xi_\epsilon\sim g^{-1}\ln^2(g/|\epsilon|)$. 
The existence of two length scales was interpreted by
Fisher~\cite{Fis94} as a consequence of fluctuations 
(see also \cite{BalFis97}). The ``typical''
localisation length $\widetilde\xi_\epsilon$ characterises the decay
of a typical wave function (more precisely $\widetilde\xi_\epsilon$ is
related to the average of the {\it logarithm} of the wave function) 
and the ``average'' localisation length $\xi_\epsilon$
controls the decay of the {\it average} correlation functions. 
Since we have considered an {\it average} quantity
$\mean{\varrho(\epsilon;L)}$,  the fact that we have extracted the
scale $\xi_\epsilon$ is consistent with Fisher's argument. 

Several arguments support the existence of a delocalisation transition
at $\epsilon\to0$. Yet they are of quite different nature and
it is interesting to provide a brief review.
The first four arguments are bulk properties of the model.
({\it i}) The Lyapunov exponent vanishes at low energies:
$\gamma(\epsilon)=1/\widetilde\xi_\epsilon\to0$ for
$\epsilon\to0$~\cite{BouComGeoLeD90}.  
({\it ii}) The calculation of the average Green's
function~\cite{BouComGeoLeD90,BalFis97,SteFabGog98}. 
({\it iii}) The DC conductivity of the model was computed at the Dirac
point in Ref.~\cite{GogMel77,Gog82} and was found to be
finite\footnote{
  In the localised phase in 1d, the AC conductivity vanishes at small
  frequency according to Mott's law
  $\re\sigma(\omega)\sim\omega^2\ln^2\omega$~\cite{AntPasSly81}. 
}.
({\it iv}) The statistical properties of the zero mode wave function
\cite{BroKre95,SheTsv98,ComTex98} indicate long range power law
correlations.
Another two arguments are obtained from scattering analysis~:
({\it v}) The distribution of the transmission probability through a
finite slab of length $L$ at zero energy was obtained in
Ref.~\cite{SteCheFabGog99}~; in particular the average transmission
decreases like $1/\sqrt{L}$, that is slower than the behaviour $1/L$
for a quasi-1d weakly-disordered conducting wire.
({\it vi}) 
The time delay distribution presents a log-normal
distribution at zero energy~\cite{SteCheFabGog99,Tex99}
$P_L(\tau\to\infty)\sim\frac1{\tau\sqrt{L}}\exp-\frac1{8gL}\ln^2(g\tau)$,
with moments diverging with the length of the disordered region~\footnote{
  This suggests that a particle of energy $\epsilon\approx0$ injected
  in a disordered region of size $L$ will spread over the full
  interval and remain trapped a very long time depending on $L$~;
  whereas in a strongly localised phase the particle would only
  explore a typical size $\widetilde{\xi}_\epsilon$ what is reflected in the
  fact that the time delay distribution reaches a limit distribution
  $\lim_{L\to\infty}P_L(\tau)=P_\infty(\tau)$~\cite{TexCom99}.  }~\cite{Tex99}
$\mean{\tau(0)}=2L$ and
$\mean{\tau(0)^n}\sim{}g^{-n}\EXP{2n^2gL}$.
Finally, two arguments consider the problem on a finite
interval taking into account boundary conditions~: 
({\it vii}) The study of extreme value
statistics of energy levels reveals spectral correlations for
$\epsilon\to0$~\cite{Tex00}, whereas a localised phase is
characterised by absence of level correlations~\cite{Mol81}.
({\it viii}) Finally the present work has shown the influence of the
boundary on $\mean{\varrho(\epsilon\to0;L)}$ due to delocalisation.

It remains an interesting issue to go beyond this analysis of the
localisation characterised by the two lengths $\widetilde\xi_\epsilon$
and $\xi_\epsilon$, and comprehend better the role of fluctuations, for
example for the DoS~$\varrho(\epsilon;L)$.  
Another challenging issue would be to extend our results to two
dimensions. Since our method is specific to one-dimensional systems,
such an extansion would necessitate the development of other methods.
This might be of interest due to the recent revival of two-dimensional 
random Dirac Hamiltonian physics, motivated by the study of Graphene.

\section*{Acknowledgements}
It is our pleasure to acknowledge Alain Comtet 
for stimulating discussions, and in particular for pointing
out the relation discussed in appendix~\ref{app:alain}.
Moreover, we thank Pierre Le Doussal for discussions. 

\appendix

\section{Extreme value spectral statistics\label{app:pin}}
We give the first distributions introduced in the text.
We compute the integral (\ref{DefPin}) using the residue theorem.
This requires the expansion of the denominator near the pole 
$\kappa_n=-\frac{\pi^2}{4}(2n+1)^2$, $n\in\NN$~:
$\cosh\sqrt{q}=\frac{(-1)^n}{\pi(2n+1)}(q-\kappa_n)\big[
  1 + \frac{q-\kappa_n}{-4\kappa_n}
  + \frac{12+4\kappa_n}{6}\big(\frac{q-\kappa_n}{-4\kappa_n}\big)^2
  + \frac{10+4\kappa_n}{2}\big(\frac{q-\kappa_n}{-4\kappa_n}\big)^3
  +\cdots
\big]$.
Some algebra gives~:
\begin{align}
  \label{pi1}
  \pi_1(y) &= \sum_{n=0}^\infty (-1)^n \pi (2n+1)\, \EXP{ -\frac{\pi^2}{4}(2n+1)^2 y }
  \\
  \label{pi1b}
  & = \frac{ 1  }{\sqrt{\pi}\,y^{3/2}}
  \sum_{n=0}^\infty (-1)^n (2n+1)\, \EXP{ -(2n+1)^2/4y }\\
  \label{pi2}
    \pi_2(y) &=\sum_{n=0}^\infty
  \left[\pi^2(2n+1)^2y - 2\right] \EXP{ -\frac{\pi^2}{4}(2n+1)^2 y }
  \\
  \label{pi2b}
  &=\frac{4}{\sqrt{\pi}\,y^{3/2}}
  \sum_{n=0}^\infty(-1)^{n+1} n^2\EXP{-n^2/y}\\
  \label{pi3}
  \pi_3(y) &=\sum_{n=0}^\infty  (-1)^n \pi (2n+1)
                     \bigg[ \frac{\pi^2(2n+1)^2}{2} y^2 - 3  y
                                 +\frac12  \bigg]
                       \EXP{-\frac{\pi^2}{4}(2n+1)^2y} 
    \\
  \label{pi3b}
    &=\frac{ 1  }{2\sqrt{\pi}\,y^{3/2}}
  \sum_{n=0}^\infty (-1)^{n+1} (2n+1)\, \left[ (2n+1)^2 -1 \right] \,\EXP{ -(2n+1)^2/4y }\\
  \label{pi4}
    \pi_4(y) &=\sum_{n=0}^\infty
                       \bigg[\frac{\pi^4(2n+1)^4}{6} y^3 - 2\pi^2(2n+1)^2 y^2 
+2\left(\frac{\pi^2(2n+1)^2}{3}+1\right) y- \frac43\bigg]
\EXP{-\frac{\pi^2}{4}(2n+1)^2y} 
\\ 
  \label{pi4b}
&= \frac{8}{3\sqrt{\pi}\,y^{3/2}}\sum_{n=0}^\infty (-1)^n n^2(n^2-1)\EXP{-n^2/y}
\end{align}
(the distribution $\pi_1(y)$ was already obtained in
Ref.~\cite{FisYou98} where the energy $\epsilon_1$
is interpreted as the gap of a spin chain. It is also related to the
distribution of the random bond lengths in the real space
renormalisation group procedure \cite{Fis94,Fis95,LeDMonFis99,LeD09}~;
$\pi_2(y)$ and $\pi_4(y)$ are given in Ref.~\cite{Tex00}).

Expressions (\ref{pi1b},\ref{pi2b},\ref{pi3b},\ref{pi4b}) are obtained from
eqs.~(\ref{pi1},\ref{pi2},\ref{pi3},\ref{pi4}) by using the Poisson formulae 
given in the next appendix and 
$\pi_3(y)=(-2y^2\frac{\D}{\D y}-3y+\frac12)\pi_1(y)$,
$\pi_2(y)=-2(2y\frac{\D}{\D y}+1)\sum_{n=0}^\infty\EXP{ -\frac{\pi^2}{4}(2n+1)^2 y }$ and
$\pi_4(y)=(\frac83y^3\frac{\D^2}{\D y^2}+8(y^2-\frac{y}3)\frac{\D}{\D y}+2y-\frac43)\sum_{n=0}^\infty\EXP{ -\frac{\pi^2}{4}(2n+1)^2 y }$.

\subsection{Generating function}
We propose now a method that allows for a systematic determination of
the distributions $\pi_n(y)$. Let us introduce the generating function
\begin{equation}
  \mathcal{G}(z,y) = \sum_{n=1}^\infty z^n \pi_n(y)
  = z \int_\mathcal{B}\frac{\D q}{2\I\pi}\frac{\EXP{q y}}{\cosh\sqrt{q}-z} 
\end{equation}
where summation was performed in the convergence radius $|z|<1$.
The function $\acos$ is single valued in the convergence disk, therefore
we can write the poles of the integrand as 
$q_n=-(n\pi+\acos{z})^2$ with $n\in\NN$.
We compute the residues by using
$\frac1{2\sqrt{q_n}}\sinh\sqrt{q_n}=(-1)^n\frac{\sqrt{1-z^2}}{2(n\pi+\acos{z})}$
(it is helpful to notice that $\acos(x\pm\I0^+)=\acos(x)\mp\I0^+$ for
$x$ real in $[-1,+1]$), whence
\begin{equation}
  \mathcal{G}(z,y) = \frac{z}{y}\derivp{}{z}
  \sum_{n=0}^\infty (-1)^n\,\EXP{-(n\pi+\acos{z})^2y}
\end{equation}

In order to apply the Poisson summation formula \eqref{Poisson1} and
get the generating function of the distributions with odd indices, 
we exploit the symmetry with respect to a change in sign of the argument 
$\sum_{n=0}^\infty(-1)^n\,\EXP{-(n\pi+\acos{z})^2y}=-\sum_{n=-\infty}^{-1}(-1)^n\,\EXP{-(n\pi+\acos(-z))^2y}$~:
\begin{align}
  \mathcal{O}(z,y)&=\frac{\mathcal{G}(z,y)-\mathcal{G}(-z,y)}{2}
  =\frac{z}{2y}\derivp{}{z}
  \sum_{n=-\infty}^{+\infty} (-1)^n\,\EXP{-(n\pi+\acos{z})^2y} \nonumber \\
  &=\frac{z}{y} \derivp{}{z}\, \frac1{\sqrt{\pi y}}
  \sum_{n=0}^\infty T_{2n+1}(z)\,\EXP{-\frac{(n+1/2)^2}y}.
\label{eqn:genfunc}
\end{align}
Here the $T_n(z)$ are the Chebychev polynomials of the first kind
$T_n(x)=\cos(n\acos(x))$. They may be rewritten as
\begin{equation}
  T_n(x)=\frac{n}2 \sum_{k=0}^{\lfloor n/2\rfloor}
  (-1)^k \frac{(n-k-1)!}{k!(n-2k)!}\,(2x)^{n-2k}
\end{equation}
where $\lfloor x\rfloor$ is the integer part. Upon insertion into \eqref{eqn:genfunc} we obtain
\begin{equation}
  \frac1z\mathcal{O}(z,y) = \sum_{n=0}^\infty z^{2n}\,\pi_{2n+1}(y)
  = \frac1{\sqrt\pi\,y^{3/2}} \sum_{n=0}^\infty
  (-1)^n(2n+1)\EXP{-\frac{(n+1/2)^2}y}
  \sum_{k=0}^n(-1)^k \left(
   \begin{array}{c}
   	  n+k\\
	  n-k
   \end{array}
   \right)\,(2z)^{2k}.
\end{equation}
Using that $\left(\begin{array}{c} n \\ p \end{array}\right)=0$ for
$p>n$, we may relax the constraint on the summation with respect to
$k$ and extract \eqref{eqn:piodd}. Note that eq.~\eqref{eqn:piodd}
allows to recover \eqref{pi1b} and \eqref{pi3b}. 

In order to
evaluate  $\pi_n(y)$ with $n=2k$ recall that we deal with a
distribution of $n$ positive {\it i.i.d.} random variables. It follows
that we may obtain $\pi_{2k}(y)$ by convolution of $\pi_{2k-1}(y)$ and
$\pi_{1}(y)$~: 
$\pi_{2k}(y)= \int_0^y\diff x\, \pi_{2k-1}(x)\pi_1(y-x)$. 
The integration is fairly cumbersome but one may verify that it yields \eqref{eqn:pieven}.
This completes the computation of $\pi_{n}(y)$ for all positive integers 
$n$ as announced in the main text. One may verify that the summation of all
$\pi_{2k}(y)$ (or $\pi_{2k+1}(y)$) yields the densities given
by~eq.~\eqref{RES2}.  

\section{Two useful Poisson formulae\label{app:poisson}}
Let us start by recalling the well-known Poisson formula
$\sum_{n\in\ZZ} f( n ) = \sum_{n\in\ZZ} \hat f(2\pi n)$
for any function $f(x)$ defined on $\RR$, with $\hat{f}(k)=\int_{\RR}\D{}x\,\EXP{-\I{}kx}f(x)$ its Fourier transform.
Applying this formula we obtain
\begin{equation}
  \label{Poisson1}
  \sum_{n\in\ZZ}\EXP{2\I\pi n\eta}\, \EXP{-\pi^2(n+\alpha)^2y} 
  = \frac{1}{\sqrt{\pi y}} 
  \sum_{n\in\ZZ}\EXP{2\I\pi(n-\eta)\alpha}\,\EXP{-\frac{(n-\eta)^2}y}
  \:,
\end{equation}
used for $\pi_n(y)$ with even indices (set $\eta=0$ and $\alpha=1/2$)
and for the DoS (set $\eta=0$ or $\eta=1/2$ and $\alpha=0$).
For the distributions $\pi_n(y)$ with odd indices we need (with
$\eta=\alpha=1/2$) 
\begin{equation}
  \label{Poisson2}
  \sum_n  (n+\alpha)\, \EXP{2\I\pi n\eta}\,\EXP{-\pi^2(n+\alpha)^2y}
  =\frac{1}{\I(\pi y)^{3/2}}
  \sum_n (n-\eta)\, \EXP{ 2\I\pi(n-\eta)\alpha}\,\EXP{-\frac{(n-\eta)^2}y }
  \:.
\end{equation}

\section{A probabilistic interpretation of the
  result for type (S) boundary conditions\label{app:alain}} 
It is worth pointing that the function $\mathcal{D}_S(y)$ can be 
interpreted as the integrated distribution of the maximum of a
Brownian excursion. 
Let us denote by $(x(t),\,0\leq t\leq1)$ such an
excursion. We establish a relation between the distribution of the
maximum of such an excursion and the function $\mathcal{D}_S(y)$.

\subsection{Maximal height of a Brownian excursion} 
Let us consider the distribution of the maximum $M$ of a Brownian
excursion $(x(t),\,0\leq{}t\leq1)$ (a Brownian bridge constraint to be
positive). It can be written as a ratio of two path integrals~:
\begin{equation}
  \mathrm{Proba}\left[ x(t)\leq M  \right]
  =\lim_{x_0\to0^+}
  \frac{\displaystyle 
        \int_{x(0)=x_0}^{x(1)=x_0}\hspace{-0.5cm}\mathcal{D}x(t)\:
         \EXP{-\frac12\int_0^1\D t\,{\dot{x}}^2}
         \prod_{t=0}^1\theta(x(t))\theta(M-x(t))
          } {\displaystyle 
          \int_{x(0)=x_0}^{x(1)=x_0}\hspace{-0.5cm}\mathcal{D}x(t)\:
         \EXP{-\frac12\int_0^1\D t\,{\dot{x}}^2}
         \prod_{t=0}^1\theta(x(t))
         }
  \:,
\end{equation}
where $x_0>0$ is a regulator.
The ratio of path integrals may be rewritten as 
\begin{equation}
  \mathrm{Proba}\left[ x(t)\leq M  \right]
  = \lim_{x_0\to0^+}
   \frac{ \bra{x_0} \EXP{-H_1} \ket{x_0} }{ \bra{x_0} \EXP{-H_0} \ket{x_0} }
  \:,
\end{equation}
where $H_0=-\frac12\frac{\D^2}{\D x^2}$ acts on functions defined on
$\RR^+$ satisfying Dirichlet boundary condition at $x=0$ and
$H_1=-\frac12\frac{\D^2}{\D x^2}$ acts on functions defined on $[0,M]$
with Dirichlet boundary conditions.  It follows that~\cite{Chu76,BiaPitYor01} 
\begin{equation}
  \label{MaxBrEx}
  \mathrm{Proba}\left[ x(t)\leq M  \right]
  = \sqrt{\frac2\pi}\,\left(\frac\pi{M}\right)^3\:
  \sum_{n=1}^\infty n^2 \,
  \EXP{ -\frac12 \left(\frac{n\pi}{M}\right)^2 }
  \:.
\end{equation}

\subsection{First-exit time} 
We consider a Brownian motion on $[0,1/\sqrt2]$ with a reflecting
boundary condition at $x=1/\sqrt2$.
We denote by $\tau$ the time at which $x(t)$ hits $x=0$ for the
first time, starting from $x=1/\sqrt2$, and $\tau_x$ the time needed
to reach $x=0$, starting from $x$.
$h(x,q)=\mean{\EXP{-q\tau_x}}$ obeys the BFPE
$\frac12\frac{\D^2}{\D x^2}h(x,q)=q\,h(x,q)$ with boundary conditions 
$h(0,q)=1$ and $\partial_xh(x,q)|_{1/\sqrt2}=0$~\cite{Gar89}. We easily find~:
\begin{equation}
  h(x,q)=\frac{\cosh\sqrt{2q}(x-1/\sqrt2)}{\cosh\sqrt{q}}
  \:,
\end{equation}
therefore distribution of time $\tau$ is given by inverse Laplace transform of 
\begin{equation}
  \mean{\EXP{-q\tau}} = \frac1{\cosh\sqrt{q}}
  \:.
\end{equation}
We introduce the sum of $n$ {\it i.i.d} such variables~:
$y=\tau_1+\cdots+\tau_n$. The distribution of this variable was
introduced in the text
$
  \pi_n(y) =  \int_{-\I\infty}^{+\I\infty}\frac{\D q}{2\I\pi}\,
  \frac{\EXP{q\, y}}{\cosh^{n}\sqrt{q}}
$,
where we have shown that
$\mathcal{D}_S(y)=\sum_{n=1}^\infty\pi_{2n}(y)=\frac4{\sqrt\pi\,y^{3/2}}\sum_{n=1}^\infty{}n^2\,\EXP{-n^2/y}$.  
Comparing with \eqref{MaxBrEx} we seen that the sum of these
distributions coincides with the cumulative distribution of the
Brownian excursion~\footnote{ 
  The fact that this sum
  can be interpreted as a cumulative distribution is related to the
  fact that the average density of the variables $\tau_n$'s is unity.
}
\begin{equation}
  \mathcal{D}_S(y) 
  = \mathrm{Proba}\left[ x(t)\leq M=\pi\sqrt{y/2}  \right]
  \:.
\end{equation}
It would be interesting to know whether this remark is purely
accidental or not.

\end{document}